\documentclass[preprint,superscriptaddress]{revtex4-1}
\usepackage{graphicx}
\usepackage{amssymb}

\begin{document}

\title{Spin-split bands cause the indirect band gap of (CH$_3$NH$_3$)PbI$_3$: Experimental evidence from circular photogalvanic effect}

\author{Daniel Niesner$^*$\thanks{present address: Columbia University, New York City. Email: dn2348@columbia.edu}}
\affiliation{Lehrstuhl f\"ur Festk\"orperphysik, Friedrich-Alexander-Universit\"at Erlangen-N\"urnberg (FAU), Staudtstr.~7, 91058~Erlangen, Germany}
\author{Martin Hauck}
\affiliation{Lehrstuhl f\"ur Angewandte Physik, Friedrich-Alexander-Universit\"at Erlangen-N\"urnberg (FAU), Staudtstr.~7, 91058~Erlangen, Germany}
\author{Shreetu Shrestha}
\affiliation{Institute of Materials for Electronics and Energy Technology~(I-MEET), Department of Materials Science and Engineering, Friedrich-Alexander-Universit\"at Erlangen-N\"urnberg (FAU), Martensstrasse~7, 91058~Erlangen, Germany}
\author{Ievgen Levchuk}
\affiliation{Institute of Materials for Electronics and Energy Technology~(I-MEET), Department of Materials Science and Engineering, Friedrich-Alexander-Universit\"at Erlangen-N\"urnberg (FAU), Martensstrasse~7, 91058~Erlangen, Germany} 
\author{Gebhard J. Matt}
\affiliation{Institute of Materials for Electronics and Energy Technology~(I-MEET), Department of Materials Science and Engineering, Friedrich-Alexander-Universit\"at Erlangen-N\"urnberg (FAU), Martensstrasse~7, 91058~Erlangen, Germany}
\author{Andres Osvet}
\affiliation{Institute of Materials for Electronics and Energy Technology~(I-MEET), Department of Materials Science and Engineering, Friedrich-Alexander-Universit\"at Erlangen-N\"urnberg (FAU), Martensstrasse~7, 91058~Erlangen, Germany}
\author{Miroslaw Batentschuk}
\affiliation{Institute of Materials for Electronics and Energy Technology~(I-MEET), Department of Materials Science and Engineering, Friedrich-Alexander-Universit\"at Erlangen-N\"urnberg (FAU), Martensstrasse~7, 91058~Erlangen, Germany}
\author{Christoph Brabec}
\affiliation{Institute of Materials for Electronics and Energy Technology~(I-MEET), Department of Materials Science and Engineering, Friedrich-Alexander-Universit\"at Erlangen-N\"urnberg (FAU), Martensstrasse~7, 91058~Erlangen, Germany}
\affiliation{Bavarian Center for Applied Energy Research (ZAE Bayern), Haberstrasse 2a, 91058 Erlangen, Germany}
\author{Heiko B. Weber}
\affiliation{Lehrstuhl f\"ur Angewandte Physik, Friedrich-Alexander-Universit\"at Erlangen-N\"urnberg (FAU), Staudtstr.~7, 91058~Erlangen, Germany}
\author{Thomas Fauster}
\affiliation{Lehrstuhl f\"ur Festk\"orperphysik, Friedrich-Alexander-Universit\"at Erlangen-N\"urnberg (FAU), Staudtstr.~7, 91058~Erlangen, Germany}

%\keywords{Organic-inorganic perovskite, photocurrent, circular photogalvanic effect, photoluminescence, indirect gap, CH3NH3PbI3}

\date{\today}

\begin{abstract}

Long carrier lifetimes and diffusion lengths form the basis for the successful application of the organic-inorganic perovskite (CH$_3$NH$_3$)PbI$_3$ in solar cells and lasers. The mechanism behind the long carrier lifetimes is still not completely understood. Spin-split bands and a resulting indirect band gap have been proposed by theory. Using near band-gap left-handed and right-handed circularly polarized light we induce photo\-currents of opposite directions in a single-crystal (CH$_3$NH$_3$)PbI$_3$ device at low temperature (4~K). The phenomenom is known as the circular photo\-galvanic effect and gives direct evidence for  photo\-transport in spin-split bands. Simultaneous photoluminecence measurements show that the onset of the photo\-current is below the optical band gap.  The results prove that an indirect band gap exists in (CH$_3$NH$_3$)PbI$_3$ with broken inversion symmetry as a result of spin-splittings in the band structure. This information is essential for understanding the photo\-physical properties of organic-inorganic perovskites and finding lead-free alternatives. Furthermore, the optically driven spin currents in (CH$_3$NH$_3$)PbI$_3$ make it a candidate material for spin\-tronics applications.

\end{abstract}

\maketitle

Organic-inorganic perovskite semiconductors (OIPS) show remarkable potential for applications in highly efficient thin-film solar cells~\cite{Yang2015, saliba2016, stranks2015} and nanolasers~\cite{zhu2015}. Unusually long carrier lifetimes~\cite{Bi2016, xu2016iodomethane} and diffusion lengths~\cite{stranks2013electron, dong2015} form the basis of their exceptional performance in opto\-electronic devices. Strong spin-orbit coupling due to the constituting heavy elements~\cite{even2013, brivio2014, quarti2014, demchenko2016} and a resulting slightly indirect band gap have been proposed as origin of the observed long carrier lifetimes~\cite{zheng2015, etienne2016, azarhoosh2016}. The direct-indirect character of the band gap of (CH$_3$NH$_3$)PbI$_3$ was recently evidenced experimentally~\cite{hutter2016,wang2016}. Direct experimental evidence for spin-orbit coupling as the origin of the indirect band gap, however, is still missing to the best of our knowledge.

\begin{figure}[t]\bf
\center
\includegraphics[width=0.5\columnwidth]{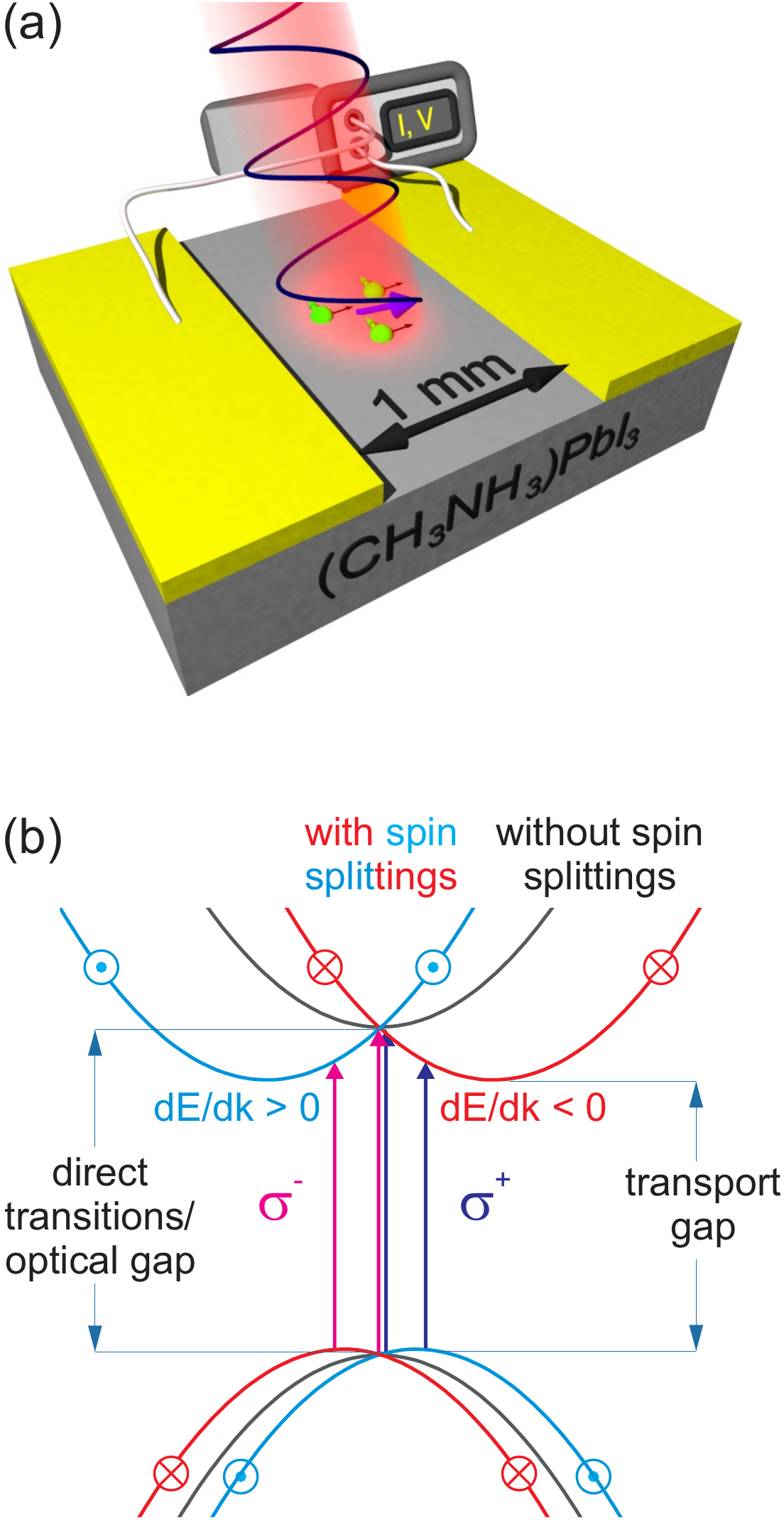}
\caption{\textbf{Circular photogalvanic effect: Experimental setup and schematic illustration in momentum space.} \rm (a) The (CH$_3$NH$_3$)PbI$_3$ single crystal with gold contacts (channel width: 1~mm) is illuminated by monochromatic cw light of tunable photon energy close to the band gap. The angle of incidence is 31$^\circ$ with respect to the surface normal. %is required for the circular photogalvanic effect to be observed~\cite{Ganichev2001}. 
(b) Rashba and Dresselhaus-type spin splittings lift the spin-degeneracy of electronic bands. Excitation with left- ($\sigma^-$) and right-handed ($\sigma^+$) circularly polarized light creates photocarriers on opposite branches of the band structure. Since the associated group velocities d$E$/d$k$ differ, a spin current is induced as indicated in (a).}	 
\label{fig1}
\end{figure}

To gain insight into the mechanism giving rise to the indirect gap of (CH$_3$NH$_3$)PbI$_3$, we excite photo\-currents with left-handed and right-handed circularly polarized light as illustrated in Fig.~\ref{fig1}~(a). In the absence of spin-orbit coupling the direction of the excited photocurrent does not depend on the helicity of the incoming light. The spin structure of the electronic band structure causes differences in the optical transition matrix elements as illustrated in Fig.~\ref{fig1}~(b). For opposite helicity of the light the group velocity of carriers is reversed and spin-polarized currents of opposite direction are induced~\cite{belinicher1978, ganichev2014}. They enhance or reduce the overall photo\-current, respectively. The effect is known as the circular photogalvanic effect. It has been observed experimentally in GaAs/AlGaAs quantum well structures~\cite{lechner2011}, in wurtzite semiconductors such as ZnO~\cite{zhang2010} and GaN~\cite{Weber2005}, in transition-metal dichalcogenides~\cite{yuan2014}, and in the topological insulator Bi$_2$Se$_3$~\cite{mciver2012}. A circular photo\-galvanic effect of measurable magnitude has been predicted~\cite{li2016circular} for (CH$_3$NH$_3$)PbI$_3$. Previously, circular dichroism has been found in optical~\cite{giovanni2015} and electron spectroscopy~\cite{niesner2016} experiments on OIPS. A circular photo\-galvanic effect is hence expected if coherent spin transport takes place on a length scale large enough for spin-polarized currents to be driven through a device.

\begin{figure}\bf
\center\includegraphics[width=0.5\columnwidth]{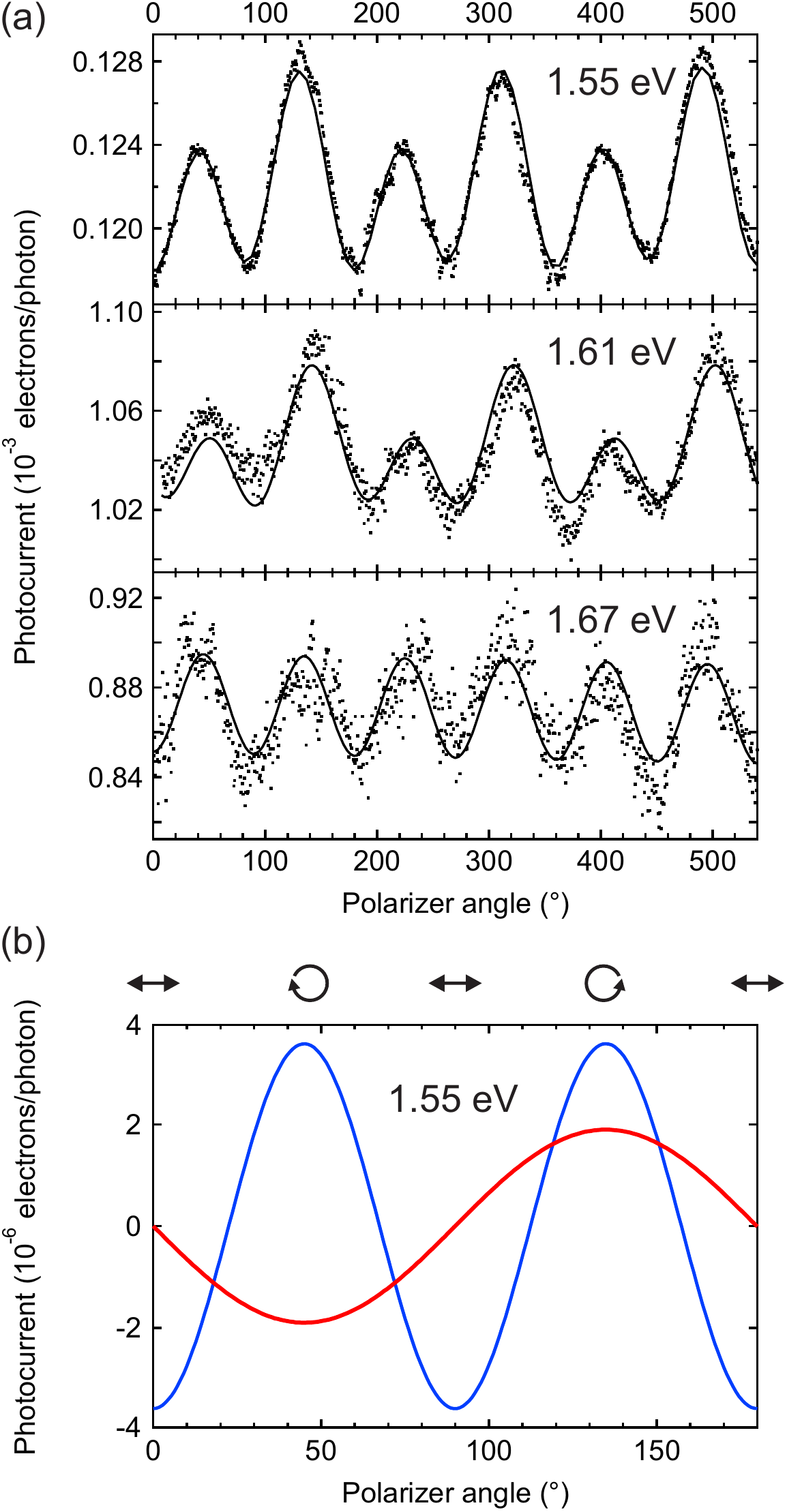}
\caption{\textbf{Polarization dependence of photocurrent.} \rm (a) The photocurrent induced with laser light of photon energies (1.55~eV and 1.61~eV) below the band gap (1.64~eV) shows a periodicity of 180$^\circ$ with respect to the angle of the $\lambda$/4 waveplate, which changes to 90$^\circ$ for above band gap excitation (1.67~eV). Photocurrents are normalized to the incoming photon flux. (b) shows results of a fit to the data in (a). The phase shift between the cosine components with 90$^\circ$ periodicity (red) and 180$^\circ$ periodicity (blue) is 90$^\circ$. The latter hence represents a circular photogalvanic signal with opposite directions of the currents induced by left-handed and right-handed circularly polarized light.}	 
\label{fig3}
\end{figure}

Results of polarization-dependent photo\-current measurements performed on single-crystal (CH$_3$NH$_3$)PbI$_3$  are shown in Fig.~\ref{fig3}~(a) for different excitation photon energies. The sample temperature is 4~K. To control the polarization of the incident light a zero-order $\lambda$/4 plate is introduced in the excitation pathway. The polarizing waveplate is rotated while the photocurrent is measured. The angle of the $\lambda$/4 plate is given along the horizontal axis in Fig.~\ref{fig3}~(a). At all wavelengths, a variation of the photocurrent is observed due to changing contributions of $p$ and $s$-polarized light fields to the excitation. This signal has a periodicity of 90$^\circ$ in the angle of the waveplate. At an angle of $n \cdot 90^\circ$ ($n \in \mathbb{N}_0$) the light is $p$-polarized and the photocurrent has a minimum. Local maxima occur at  45$^\circ$ (135$^\circ$), where the light is circularly polarized and the component of $s$-polarized light is the largest. The variations may result from differences in reflectivity at the surface of the OIPS due to changing contributions of $p$ and $s$-polarized light, from anisotropies in absorption along the crystalline directions associated with $s$ and $p$ polarization, and from a linear photogalvanic effect~\cite{Ganichev2003}. Since differences in the photo\-currents excited with $s$-polarized light and circularly polarized light with $p$-components can occur in any material and their interpretation is complex, they will not be in the focus of our discussion. However, it is worth noting that a linear photo\-galvanic effect necessarily goes hand in hand with the circular photo\-galvanic effect~\cite{Ganichev2003, mciver2012}.

\begin{figure}[t]\bf
\center\includegraphics[width=0.8\columnwidth]{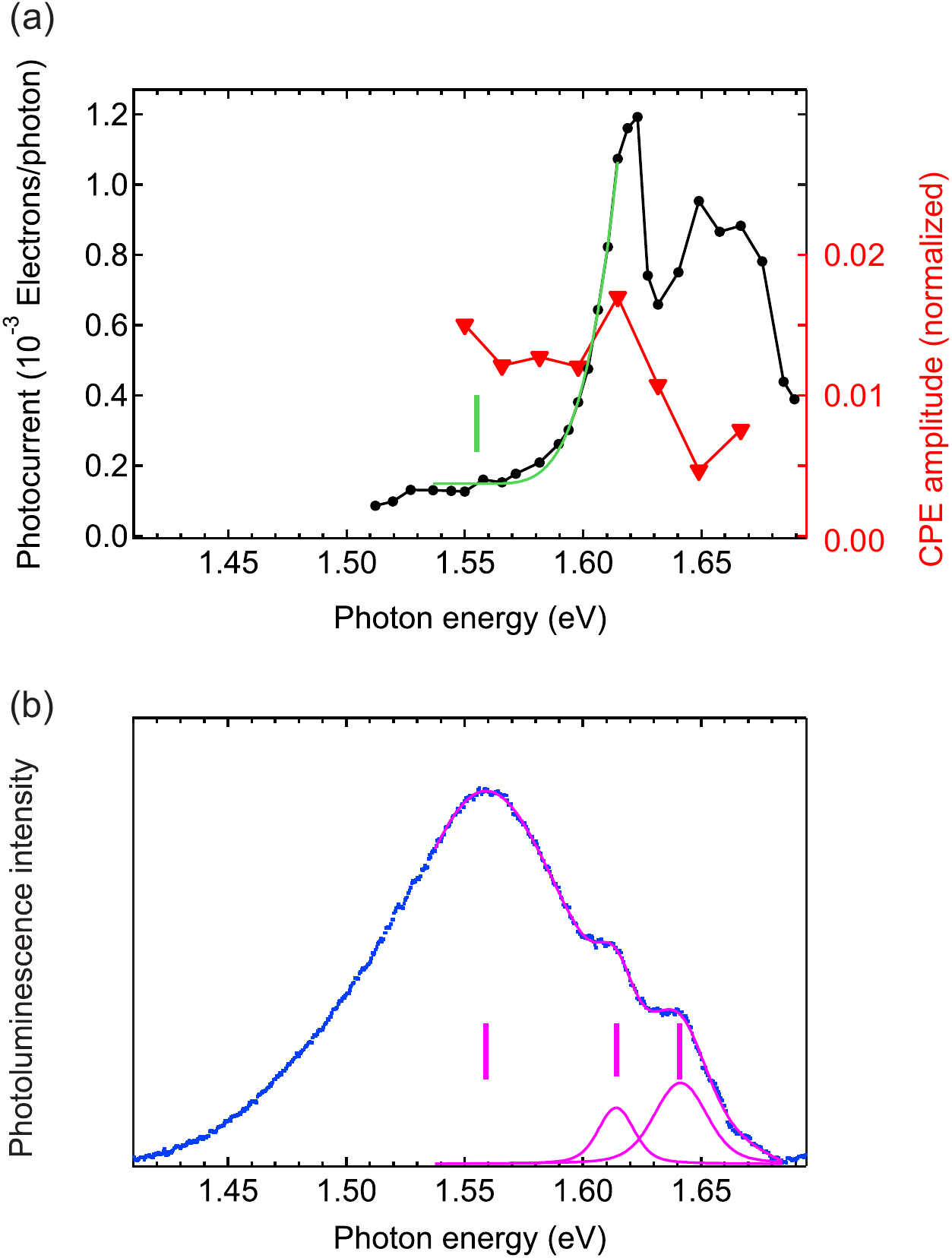}%\includegraphics[width=0.8\columnwidth,clip,trim=0 0 0 217]{abb2.eps}

%\center\includegraphics[width=0.8\columnwidth,clip,trim=0 242 0 0]{abb2.eps}

\caption{\textbf{Transport gap and optical gap of (CH$_3$NH$_3$)PbI$_3$.} \rm (a) Photocurrent excited by monochromatic laser light (black symbols). The onset of the photocurrent (tick mark) is extracted by a power law fit (green curve) yielding a fourth power dependence. 
The amplitude of the circular photo\-galvanic effect (CPE amplitude) normalized to the average photocurrent extracted from Fig.~\ref{fig3}~(a) is shown by red symbols. The data points are connected for clarity. (b) The photoluminescence spectrum (blue symbols) is fitted (magenta) by a sum of three Voigtian peaks. Peak positions are given by tick marks. 
}	 
\label{fig2}
\end{figure}

An additional modulation of the photocurrent induced by the light polarization is clearly observed upon excitation at 1.55~eV and 1.61~eV photon energy. This signal has a periodicity of 180$^\circ$ in the angle of the waveplate, resulting in different photocurrents at $45^\circ + n \cdot 180^\circ$  and $135^\circ + n \cdot 180^\circ$. As these angles correspond to left-handed %(as seen from the source) 
and right-handed circularly polarized light, the differences represent the circular photogalvanic effect. To extract the contribution of the circular photogalvanic effect to the photocurrent, we fit the data with a sum of two cosine functions. The two components are shown individually in Fig.~\ref{fig3}~(b). The effect of linear polarization is given by the blue curve. The signal arising from the circular photogalvanic effect, indicated by the red curve, is phase-shifted by 90$^\circ$. Its contribution to the photocurrent vanishes whenever the light is linearly polarized ($n \cdot 90^\circ$). For left-handed (45$^\circ$) and  right-handed (135$^\circ$) circular polarization, in contrast, it switches sign. The reversal of the photo\-currents as the helicity of the excitation light is switched is characteristic for materials with spin-split band structures~\cite{Weber2005,zhang2010,lechner2011,mciver2012,yuan2014}. It implies that spin currents are driven by photo\-excitation with circularly polarized light, as indicated in Fig.~\ref{fig1}~(b). Light of different helicity couples to opposite branches of the spin-split band structure in $k$-space. Since the opposite branches do not only carry electrons of opposite spin orientation, but also of reversed group velocity d$E$/d$k$, spin-polarized currents are induced along opposite directions. We observe a modification of the overall photocurrent by$~\pm1.5$\%. The amplitude of the circular photogalvanic effect relative to the average photocurrent is given by red symbols in Fig.~\ref{fig2} (a). 

The photocurrent (normalized to electrons/photon) is shown as black dots connected by lines to guide the eye in Fig.~\ref{fig2}(a). The onset of the photo\-current is well described by a fourth-power dependence on energy starting at $1.56\pm0.01$~eV\@. The large exponent can be understood as the result of an indirect band gap in combination with a low density of states of OIPS at the band edges~\cite{endres2016, niesner2016}. A small current flows for photon energies below the onset because of the applied bias voltage. Note that the circular photogalvanic effect sets in right at the onset of the photocurrent. The photocurrent reaches its maximum at 1.62~eV photon energy which may be taken as an estimate for the direct band gap. This results in a difference between the direct and indirect gap of $60\pm15$~meV in agreement with literature~\cite{hutter2016,wang2016,savenije2014}. The circular photogalvanic effect proves that the spin-orbit coupling of the Rashba effect is responsible for the indirect band gap in (CH$_3$NH$_3$)PbI$_3$.

In order to corroborate the assignment of the direct band gap we performed \textit{in situ} photoluminescence (PL) measurements on the sample at 4~K\@.
A comparison of photocurrent excitation and PL spectra is given in Fig.~\ref{fig2}. The low-temperature PL spectrum of (CH$_3$NH$_3$)PbI$_3$ consists of a high-energy emission feature at $1.64\pm0.01$~eV, a second peak at $1.61\pm0.01$~eV, and broad low-energy continuum emission with a maximum at $1.56\pm0.01$~eV, in agreement with previous reports~\cite{dar2016,diab2016, wu2015, galkowski2016a}. The positions of the maxima are indicated by magenta tick marks in Fig.~\ref{fig2}(b). Following the literature~\cite{diab2016}, we assign the highest-energy peak at 1.64~eV to the optical band gap of orthorhombic (CH$_3$NH$_3$)PbI$_3$. The value matches the optical band gap found from magneto-absorption measurements~\cite{miyata2015}. Emission and absorption at lower photon energies have been attributed to excitons localized at defects~\cite{wu2015, diab2016} and to coexisting structural phases~\cite{galkowski2016a, dar2016}. The assignment of optical transitions to bound or free excitons is difficult based on optical spectroscopy alone. Connecting the PL spectroscopy to the transport measurements at low temperature, we find that currents are generated with photon energies as low as 1.56~eV, demonstrating that free excitons are excited at these photon energies. We attribute the photo\-current to tetragonal~\cite{phuong2016, galkowski2016a} and low-symmetry orthorhombic~\cite{dar2016} domains coexisting with the low-temperature, inversion-symmetric orthorhombic phase. The photo\-current has a maximum at $1.62$~eV which coincides with the second PL emission peak, indicating an allowed optical transition. We assign the maximum to the direct gap of the tetragonal and low-symmetry orthorhombic domains, in good agreement with the value of 1.61~eV found from magneto-absorption on tetragonal (CH$_3$NH$_3$)PbI$_3$~\cite{miyata2015, galkowski2016}. The photo\-current drops by 40\% as the photon energy exceeds the optical gap of 1.64~eV of orthorhombic (CH$_3$NH$_3$)PbI$_3$. A smaller photo\-current in orthorhombic (CH$_3$NH$_3$)PbI$_3$ than in the tetragonal phase has been reported before and assigned to less efficient generation of free excitons~\cite{hutter2016}. Note that the amplitude of the circular photogalvanic effect also drops in the energy range when the inversion symmetric low-temperature orthorhombic phase contributes to the photocurrent. 

The observed circular photogalvanic effect unambiguously identifies spin splittings in the band structure as the origin of the indirect gap.
A slightly indirect band gap by 47~meV was reported for tetragonal (CH$_3$NH$_3$)PbI$_3$ resulting in an enhanced lifetime of photo\-excited carriers as compared to the direct band-gap orthorhombic phase~\cite{hutter2016}. The lifetime enhancement was found to be absent in the low-temperature orthorhombic phase. 
While calculations point towards Rashba and Dresselhaus type spin splittings as the origin of the slightly indirect gap~\cite{even2013, brivio2014, quarti2014, demchenko2016, wang2016}, experimental evidence for this interpretation is, to the best of our knowledge, lacking.  Requirements for spin splittings are spin-orbit coupling and absence of inversion symmetry. It is worth noting that Rashba and Dresselhaus spin splittings are caused by the local environment of the atoms in the unit cell rather than by the average, long-range symmetry of the crystal~\cite{zhang2014}. A Rashba-type spin-split band structure was found in the valence band at the surface of related (CH$_3$NH$_3$)PbBr$_3$ perovskite using angle-resolved photo\-electron spectroscopy~\cite{niesner2016}.  Surfaces break inversion symmetry inherently and enhance Rashba splitting. Observation of the circular photogalvanic effect demonstrates that spin-splittings occur in the bulk of (CH$_3$NH$_3$)PbI$_3$ on a length scale relevant for carrier transport. We find a stronger effect at low photon energies than for higher ones. The corresponding transitions can be assigned to tetragonal and low-symmetry orthorhombic domains ($<1.64$~eV) and to the inversion symmetric low-temperature orthorhombic phase ($\geq 1.64$~eV), respectively. For the latter, a prominent photogalvanic effect is not expected. The former, in contrast, has a locally broken inversion symmetry at all temperatures~\cite{druzbicki2016,beecher2016}. The spin splittings in the band structure observed here at 4~K are hence expected to persist for temperatures up to room temperature, as implied also by the strong circular dichroism found in optical spectroscopy at room temperature~\cite{giovanni2015}. 

The observed difference between the optical and the transport gap of $60\pm15$~meV of tetragonal (CH$_3$NH$_3$)PbI$_3$ is large enough to pose an energetic barrier for electron-hole-pair recombination even at room temperature. Activation energies of 75~meV~\cite{savenije2014} and 47~meV~\cite{hutter2016} for radiative recombination were reported previously, in good agreement with our results. Calculations find a value of 75~meV as a result of spin-splittings in the band structure of (CH$_3$NH$_3$)PbI$_3$~\cite{azarhoosh2016}. They predict an increasing splitting with increasing temperature, in agreement with optical spectroscopic results obtained from related (CH$_3$NH$_3$)PbBr$_3$ single crystals~\cite{niesner2016optical}.

Our results clarify the mechanism behind the indirect character of the band gap of (CH$_3$NH$_3$)PbI$_3$. This helps to understand the excellent performance of OIPS in optoelectronic devices and provides a design rule for less toxic alternatives to (CH$_3$NH$_3$)PbI$_3$. Lifetime enhancements by a factor of 10 to 350 have been predicted as the result of Rashba-type spin splitting restricting optical transitions~\cite{zheng2015, etienne2016, azarhoosh2016}, making it an essential ingredient to the observed long carrier diffusion lengths. ~\cite{hutter2016}
 
Spin splittings of this magnitude do not only enhance carrier lifetimes. They also allow to optically drive spin currents~\cite{kepenekian2015, Li2016, zhang2017} in the system. %, opening the possibility to apply OIPS in opto-spintronics devices~\cite{kepenekian2015, Li2016, zhang2017}. 
The spin splitting of $60\pm 15$~meV is similar to the strongest ones found in known bulk Rashba systems, such as \mbox{BiTeX} (X~$=$~Cl~\cite{landolt2013}, Br~\cite{ogawa2014}, I~\cite{Ishizaka2011}) and GeTe(111)~\cite{liebmann2016}. In contrast to these materials, (CH$_3$NH$_3$)PbI$_3$ has a band gap in the near-infrared range making it a candidate material for opto-spintronics applications~\cite{Li2016}. Additional applications become possible if structures with a switch\-able ferro\-electric polarization can be grown~\cite{kim2014, leppert2016}, as they have been found at the surface of (CH$_3$NH$_3$)PbI$_3$~\cite{kutes2014}, where the Rashba splitting in organic-inorganic perovskite is further enhanced~\cite{niesner2016}. 

 We find a measurable circular photogalvanic effect of $\pm1.5$\% in rather large devices with a channel width of 1~mm. Significant spin currents can be expected when device dimensions are reduced to the spin transport length. From magneto-transport and magneto-optical experiments, a spin-lattice relaxation time $\tau$ of 200~ps was estimated for spin-cast (CH$_3$NH$_3$)PbI$_3$ thin films~\cite{zhang2015magnet}. Carrier diffusion coefficients in (CH$_3$NH$_3$)PbI$_3$ thin films are around $D=0.05$~cm$^2\mathrm{s}^{-1}$~\cite{stranks2013electron, guo2015spatial}, translating into a spin diffusion length $l=\sqrt{D \cdot \tau}=30$~nm. The carrier diffusion constant in single crystals is larger than in thin films by a factor of $\approx 20$~\cite{shi2015, dong2015, guo2015spatial}, and also the spin relaxation time can be expected to be enhanced. Spin diffusion lengths of hundreds of nanometers may hence be achieved in OIPC single crystals. 

%\section*{Author contributions}
%D. N., C. J. B., and T. F. developed the original idea. S. S., I. L., G. J. M., and A.~O. fabricated and characterized perovskite single-crystal devices supervised by M. B. and C.~J.~B.. D. N. and M. H. performed the photocurrent and photoluminescence experiments under supervision by H. B. W. and T. F.. D.~N., M. H., and T. F. prepared the figures. D.~N., H. B. W., and T. F. wrote the first version of the manuscript. All authors were actively involved in writing the final version of the manuscript.

%\section*{Competing financial interests}
%The authors declare no competing financial interests.

\section*{Acknowledgements}

I.~L., A.~O., S.~S., M.~B., and C.~J.~B. gratefully acknowledge financial support from the Soltech Initiative, the Excellence Cluster "Engineering of Advanced Materials" (EAM) granted to the University Erlangen-Nuremberg, and from the Energiecampus N\"urnberg. Funding from the Emerging Fields initiative "Singlet Fission" supported by Friedrich-Alexander-Universit\"at Erlangen-N\"urnberg is gratefully acknowledged by D.~N., M.~W., and T.~F.

\section*{Methods}

\paragraph*{\bf Device fabrication.}
(CH$_3$NH$_3$)PbI$_3$ single crystals were grown by the seed-solution growth method following the procedure described in Ref.~\cite{zhou2016giant}. Crystals were prepared and kept under N$_2$ atmosphere before they got contacted with 40~nm of gold at a spacing of 1~mm. Contacts are aligned with the macroscopic facets of the crystals. The crystals are mounted under ambient conditions (exposure for $\approx 2$~h) in a vacuum cryostat. Immediately after pumping the cryostat, crystals are cooled to 4~K. Keeping the crystals in vacuum at room temperature for extended periods of time ($>12$~h) results in changes in the spectra, whereas no changes are observed for the cooled crystals under prolonged (4~h) illumination with laser light, see Fig.~S1 in Supplementary Information. We demonstrated that methylamine desorbs from (CH$_3$NH$_3$)PbBr$_3$ in vacuum just above room temperature~\cite{niesner2016} and speculatively assign the degradation of (CH$_3$NH$_3$)PbI$_3$ to the same mechanism. 

\paragraph*{\bf Photocurrent and photoluminescence measurements.}
Photocurrents are excited with a Ti:Sa laser in cw mode. Laser powers are around 3~mW with a Gaussian spot radius of 0.3~mm.  The voltage applied to the device is swept between $-0.5$~V and $+0.5$~V to avoid slow changes~\cite{gottesman2014extremely, gottesman2015photoinduced} in the structure of (CH$_3$NH$_3$)PbI$_3$. No hysteresis is observed in the I(V) sweeps, see Fig.~S2 in Supplementary Information. The data are shown for a small bias voltage (0.25~V).  Results presented here are independent of the applied bias voltage for $0<\left|\mathrm{V}\right|<0.5$~V except for the amplitude of the measured photocurrents. Photo\-currents are normalized to the number of incident photons to account for variations in the laser power at different photon energies. The photo\-current increases linearly with excitation density as shown in Fig.~S3 in Supplementary Information.

For photo\-luminescence experiments the channel of the (CH$_3$NH$_3$)PbI$_3$ device was optically excited with a 532~nm cw laser between photocurrent measurementss. The PL spectra were recorded using an Ocean Optics HR 4000 spectrometer. A dielectric long pass filter (550~nm) was used to suppress scattered light from the illuminated surface from which the PL was collected.

\bibliographystyle{naturemag_noURL}
\bibliography{perovskites}

\begin{thebibliography}{10}
\expandafter\ifx\csname url\endcsname\relax
  \def\url#1{\texttt{#1}}\fi
\expandafter\ifx\csname urlprefix\endcsname\relax\def\urlprefix{URL }\fi
\providecommand{\bibinfo}[2]{#2}
\providecommand{\eprint}[2][]{\url{#2}}

\bibitem{Yang2015}
\bibinfo{author}{Yang, W.~S.} \emph{et~al.}
\newblock \bibinfo{title}{High-performance photovoltaic perovskite layers
  fabricated through intramolecular exchange}.
\newblock \emph{\bibinfo{journal}{Science}} \textbf{\bibinfo{volume}{348}},
  \bibinfo{pages}{1234--1237} (\bibinfo{year}{2015}).

\bibitem{saliba2016}
\bibinfo{author}{Saliba, M.} \emph{et~al.}
\newblock \bibinfo{title}{A molecularly engineered hole-transporting material
  for efficient perovskite solar cells}.
\newblock \emph{\bibinfo{journal}{Nat. Energy}} \textbf{\bibinfo{volume}{1}},
  \bibinfo{pages}{15017} (\bibinfo{year}{2016}).

\bibitem{stranks2015}
\bibinfo{author}{Stranks, S.~D.} \& \bibinfo{author}{Snaith, H.~J.}
\newblock \bibinfo{title}{{Metal-halide perovskites for photovoltaic and
  light-emitting devices}}.
\newblock \emph{\bibinfo{journal}{Nat. Nanotechnol.}}
  \textbf{\bibinfo{volume}{10}}, \bibinfo{pages}{391--402}
  (\bibinfo{year}{2015}).

\bibitem{zhu2015}
\bibinfo{author}{Zhu, H.} \emph{et~al.}
\newblock \bibinfo{title}{Lead halide perovskite nanowire lasers with low
  lasing thresholds and high quality factors}.
\newblock \emph{\bibinfo{journal}{Nat. Mater.}} \textbf{\bibinfo{volume}{14}},
  \bibinfo{pages}{636--642} (\bibinfo{year}{2015}).

\bibitem{Bi2016}
\bibinfo{author}{Bi, Y.} \emph{et~al.}
\newblock \bibinfo{title}{{Charge Carrier Lifetimes Exceeding 15 $\mu$s in
  Methylammonium Lead Iodide Single Crystals}}.
\newblock \emph{\bibinfo{journal}{J. Phys. Chem. Lett.}}
  \textbf{\bibinfo{volume}{7}}, \bibinfo{pages}{923} (\bibinfo{year}{2016}).

\bibitem{xu2016iodomethane}
\bibinfo{author}{Xu, W.} \emph{et~al.}
\newblock \bibinfo{title}{{Iodomethane-Mediated Organometal Halide Perovskite
  with Record Photoluminescence Lifetime}}.
\newblock \emph{\bibinfo{journal}{ACS Appl. Mater. Interfaces}}
  \textbf{\bibinfo{volume}{8}}, \bibinfo{pages}{23181--23189}
  (\bibinfo{year}{2016}).

\bibitem{stranks2013electron}
\bibinfo{author}{Stranks, S.~D.} \emph{et~al.}
\newblock \bibinfo{title}{Electron-hole diffusion lengths exceeding 1
  micrometer in an organometal trihalide perovskite absorber}.
\newblock \emph{\bibinfo{journal}{Science}} \textbf{\bibinfo{volume}{342}},
  \bibinfo{pages}{341--344} (\bibinfo{year}{2013}).

\bibitem{dong2015}
\bibinfo{author}{Dong, Q.} \emph{et~al.}
\newblock \bibinfo{title}{{Electron-hole diffusion lengths $>175~\mu$m in
  solution-grown CH$_3$NH$_3$PbI$_3$ single crystals}}.
\newblock \emph{\bibinfo{journal}{Science}} \textbf{\bibinfo{volume}{347}},
  \bibinfo{pages}{967} (\bibinfo{year}{2015}).

\bibitem{even2013}
\bibinfo{author}{Even, J.}, \bibinfo{author}{Pedesseau, L.},
  \bibinfo{author}{Jancu, J.-M.} \& \bibinfo{author}{Katan, C.}
\newblock \bibinfo{title}{{Importance of spin--orbit coupling in hybrid
  organic/inorganic perovskites for photovoltaic applications}}.
\newblock \emph{\bibinfo{journal}{J. Phys. Chem. Lett.}}
  \textbf{\bibinfo{volume}{4}}, \bibinfo{pages}{2999} (\bibinfo{year}{2013}).

\bibitem{brivio2014}
\bibinfo{author}{Brivio, F.}, \bibinfo{author}{Butler, K.~T.},
  \bibinfo{author}{Walsh, A.} \& \bibinfo{author}{van Schilfgaarde, M.}
\newblock \bibinfo{title}{{Relativistic quasiparticle self-consistent
  electronic structure of hybrid halide perovskite photovoltaic absorbers}}.
\newblock \emph{\bibinfo{journal}{Phys. Rev. B}} \textbf{\bibinfo{volume}{89}},
  \bibinfo{pages}{155204} (\bibinfo{year}{2014}).

\bibitem{quarti2014}
\bibinfo{author}{Quarti, C.}, \bibinfo{author}{Mosconi, E.} \&
  \bibinfo{author}{De~Angelis, F.}
\newblock \bibinfo{title}{{Interplay of Orientational Order and Electronic
  Structure in Methylammonium Lead Iodide: Implications for Solar Cell
  Operation}}.
\newblock \emph{\bibinfo{journal}{Chem. Mater.}} \textbf{\bibinfo{volume}{26}},
  \bibinfo{pages}{6557--6569} (\bibinfo{year}{2014}).

\bibitem{demchenko2016}
\bibinfo{author}{Demchenko, D.} \emph{et~al.}
\newblock \bibinfo{title}{{Optical properties of the organic-inorganic hybrid
  perovskite CH$_3$NH$_3$PbI$_3$: Theory and experiment}}.
\newblock \emph{\bibinfo{journal}{Phys. Rev. B}} \textbf{\bibinfo{volume}{94}},
  \bibinfo{pages}{075206} (\bibinfo{year}{2016}).

\bibitem{zheng2015}
\bibinfo{author}{Zheng, F.}, \bibinfo{author}{Tan, L.~Z.},
  \bibinfo{author}{Liu, S.} \& \bibinfo{author}{Rappe, A.~M.}
\newblock \bibinfo{title}{{Rashba Spin--Orbit Coupling Enhanced Carrier
  Lifetime in CH$_3$NH$_3$PbI$_3$}}.
\newblock \emph{\bibinfo{journal}{Nano Lett.}} \textbf{\bibinfo{volume}{15}},
  \bibinfo{pages}{7794--7800} (\bibinfo{year}{2015}).

\bibitem{etienne2016}
\bibinfo{author}{Etienne, T.}, \bibinfo{author}{Mosconi, E.} \&
  \bibinfo{author}{De~Angelis, F.}
\newblock \bibinfo{title}{{Dynamical Origin of the Rashba Effect in
  Organohalide Lead Perovskites: A Key to Suppressed Carrier Recombination in
  Perovskite Solar Cells?}}
\newblock \emph{\bibinfo{journal}{J. Phys. Chem. Lett.}}
  \textbf{\bibinfo{volume}{7}}, \bibinfo{pages}{1638} (\bibinfo{year}{2016}).

\bibitem{azarhoosh2016}
\bibinfo{author}{Azarhoosh, P.}, \bibinfo{author}{McKechnie, S.},
  \bibinfo{author}{Frost, J.~M.}, \bibinfo{author}{Walsh, A.} \&
  \bibinfo{author}{van Schilfgaarde, M.}
\newblock \bibinfo{title}{{Research Update: Relativistic origin of slow
  electron-hole recombination in hybrid halide perovskite solar cells}}.
\newblock \emph{\bibinfo{journal}{APL Mater.}} \textbf{\bibinfo{volume}{4}},
  \bibinfo{pages}{091501} (\bibinfo{year}{2016}).

\bibitem{hutter2016}
\bibinfo{author}{Hutter, E.~M.} \emph{et~al.}
\newblock \bibinfo{title}{Direct-indirect character of the bandgap in
  methylammonium lead iodide perovskite}.
\newblock \emph{\bibinfo{journal}{Nat. Mater.}} \textbf{\bibinfo{volume}{16}},
  \bibinfo{pages}{115} (\bibinfo{year}{2016}).

\bibitem{wang2016}
\bibinfo{author}{Wang, T.} \emph{et~al.}
\newblock \bibinfo{title}{Indirect to direct bandgap transition in
  methylammonium lead halide perovskite}.
\newblock \emph{\bibinfo{journal}{Energy Environ. Sci.}}
  \textbf{\bibinfo{volume}{10}}, \bibinfo{pages}{509} (\bibinfo{year}{2017}).

\bibitem{belinicher1978}
\bibinfo{author}{Belinicher, V.}
\newblock \bibinfo{title}{{Space-oscillating photocurrent in crystals without
  symmetry center}}.
\newblock \emph{\bibinfo{journal}{Phys. Lett. A}}
  \textbf{\bibinfo{volume}{66}}, \bibinfo{pages}{213--214}
  (\bibinfo{year}{1978}).

\bibitem{ganichev2014}
\bibinfo{author}{Ganichev, S.~D.} \& \bibinfo{author}{Golub, L.~E.}
\newblock \bibinfo{title}{{Interplay of Rashba/Dresselhaus spin splittings
  probed by photogalvanic spectroscopy--A review}}.
\newblock \emph{\bibinfo{journal}{Phys. Status Solidi (b)}}
  \textbf{\bibinfo{volume}{251}}, \bibinfo{pages}{1801--1823}
  (\bibinfo{year}{2014}).

\bibitem{lechner2011}
\bibinfo{author}{Lechner, V.} \emph{et~al.}
\newblock \bibinfo{title}{{Spin and orbital mechanisms of the magnetogyrotropic
  photogalvanic effects in GaAs/Al$_x$ Ga$_{1- x}$ As quantum well
  structures}}.
\newblock \emph{\bibinfo{journal}{Phys. Rev. B}} \textbf{\bibinfo{volume}{83}},
  \bibinfo{pages}{155313} (\bibinfo{year}{2011}).

\bibitem{zhang2010}
\bibinfo{author}{Zhang, Q.} \emph{et~al.}
\newblock \bibinfo{title}{{Strong circular photogalvanic effect in ZnO
  epitaxial films}}.
\newblock \emph{\bibinfo{journal}{Appl. Phys. Lett.}}
  \textbf{\bibinfo{volume}{97}}, \bibinfo{pages}{041907}
  (\bibinfo{year}{2010}).

\bibitem{Weber2005}
\bibinfo{author}{Weber, W.} \emph{et~al.}
\newblock \bibinfo{title}{{Demonstration of Rashba spin splitting in GaN-based
  heterostructures}}.
\newblock \emph{\bibinfo{journal}{Appl. Phys. Lett.}}
  \textbf{\bibinfo{volume}{87}}, \bibinfo{pages}{262106}
  (\bibinfo{year}{2005}).

\bibitem{yuan2014}
\bibinfo{author}{Yuan, H.} \emph{et~al.}
\newblock \bibinfo{title}{{Generation and electric control of
  spin--valley-coupled circular photogalvanic current in WSe$_2$}}.
\newblock \emph{\bibinfo{journal}{Nat. Nanotechnol.}}
  \textbf{\bibinfo{volume}{9}}, \bibinfo{pages}{851--857}
  (\bibinfo{year}{2014}).

\bibitem{mciver2012}
\bibinfo{author}{McIver, J.}, \bibinfo{author}{Hsieh, D.},
  \bibinfo{author}{Steinberg, H.}, \bibinfo{author}{Jarillo-Herrero, P.} \&
  \bibinfo{author}{Gedik, N.}
\newblock \bibinfo{title}{Control over topological insulator photocurrents with
  light polarization}.
\newblock \emph{\bibinfo{journal}{Nat. Nanotechnol.}}
  \textbf{\bibinfo{volume}{7}}, \bibinfo{pages}{96--100}
  (\bibinfo{year}{2012}).

\bibitem{li2016circular}
\bibinfo{author}{Li, J.} \& \bibinfo{author}{Haney, P.~M.}
\newblock \bibinfo{title}{{Circular photogalvanic effect in organometal halide
  perovskite CH$_3$NH$_3$PbI$_3$}}.
\newblock \emph{\bibinfo{journal}{Appl. Phys. Lett.}}
  \textbf{\bibinfo{volume}{109}}, \bibinfo{pages}{193903}
  (\bibinfo{year}{2016}).

\bibitem{giovanni2015}
\bibinfo{author}{Giovanni, D.} \emph{et~al.}
\newblock \bibinfo{title}{{Highly spin-polarized carrier dynamics and
  ultralarge photoinduced magnetization in CH$_3$NH$_3$PbI$_3$ perovskite thin
  films}}.
\newblock \emph{\bibinfo{journal}{Nano Lett.}} \textbf{\bibinfo{volume}{15}},
  \bibinfo{pages}{1553} (\bibinfo{year}{2015}).

\bibitem{niesner2016}
\bibinfo{author}{Niesner, D.} \emph{et~al.}
\newblock \bibinfo{title}{{Giant Rashba Splitting in CH$_3$NH$_3$PbBr$_3$
  Organic-Inorganic Perovskite}}.
\newblock \emph{\bibinfo{journal}{Phys. Rev. Lett.}}
  \textbf{\bibinfo{volume}{117}}, \bibinfo{pages}{126401}
  (\bibinfo{year}{2016}).

\bibitem{Ganichev2003}
\bibinfo{author}{Ganichev, S.~D.} \& \bibinfo{author}{Prettl, W.}
\newblock \bibinfo{title}{{Spin photocurrents in quantum well structures}}.
\newblock \emph{\bibinfo{journal}{J. Phys. Condens. Matter}}
  \textbf{\bibinfo{volume}{15}}, \bibinfo{pages}{R935} (\bibinfo{year}{2003}).

\bibitem{endres2016}
\bibinfo{author}{Endres, J.} \emph{et~al.}
\newblock \bibinfo{title}{{Valence and Conduction Band Densities of States of
  Metal Halide Perovskites: A Combined Experimental-Theoretical Study}}.
\newblock \emph{\bibinfo{journal}{J. Phys. Chem. Lett.}}
  \textbf{\bibinfo{volume}{7}}, \bibinfo{pages}{2722} (\bibinfo{year}{2016}).

\bibitem{savenije2014}
\bibinfo{author}{Savenije, T.~J.} \emph{et~al.}
\newblock \bibinfo{title}{Thermally activated exciton dissociation and
  recombination control the carrier dynamics in organometal halide perovskite}.
\newblock \emph{\bibinfo{journal}{J. Phys. Chem. Lett.}}
  \textbf{\bibinfo{volume}{5}}, \bibinfo{pages}{2189--2194}
  (\bibinfo{year}{2014}).

\bibitem{dar2016}
\bibinfo{author}{Dar, M.~I.} \emph{et~al.}
\newblock \bibinfo{title}{{Origin of unusual bandgap shift and dual emission in
  organic-inorganic lead halide perovskites}}.
\newblock \emph{\bibinfo{journal}{Sci. Adv.}} \textbf{\bibinfo{volume}{2}},
  \bibinfo{pages}{e1601156} (\bibinfo{year}{2016}).

\bibitem{diab2016}
\bibinfo{author}{Diab, H.} \emph{et~al.}
\newblock \bibinfo{title}{{Narrow Linewidth Excitonic Emission in
  Organic-Inorganic Lead Iodide Perovskite Single Crystals}}.
\newblock \emph{\bibinfo{journal}{J. Phys. Chem. Lett.}}
  \textbf{\bibinfo{volume}{7}}, \bibinfo{pages}{5093--5100}
  (\bibinfo{year}{2016}).

\bibitem{wu2015}
\bibinfo{author}{Wu, X.} \emph{et~al.}
\newblock \bibinfo{title}{Trap states in lead iodide perovskites}.
\newblock \emph{\bibinfo{journal}{J. Am. Chem. Soc.}}
  \textbf{\bibinfo{volume}{137}}, \bibinfo{pages}{2089--2096}
  (\bibinfo{year}{2015}).

\bibitem{galkowski2016a}
\bibinfo{author}{Galkowski, K.} \emph{et~al.}
\newblock \bibinfo{title}{Spatially resolved studies of the phases and
  morphology of methylammonium and formamidinium lead tri-halide perovskites}.
\newblock \emph{\bibinfo{journal}{Nanoscale}} \textbf{\bibinfo{volume}{9}},
  \bibinfo{pages}{3222} (\bibinfo{year}{2017}).

\bibitem{miyata2015}
\bibinfo{author}{Miyata, A.} \emph{et~al.}
\newblock \bibinfo{title}{Direct measurement of the exciton binding energy and
  effective masses for charge carriers in organic-inorganic tri-halide
  perovskites}.
\newblock \emph{\bibinfo{journal}{Nat. Phys.}} \textbf{\bibinfo{volume}{11}},
  \bibinfo{pages}{582} (\bibinfo{year}{2015}).

\bibitem{phuong2016}
\bibinfo{author}{Phuong, L.~Q.} \emph{et~al.}
\newblock \bibinfo{title}{{Free Carriers Versus Excitons in CH$_3$NH$_3$PbI$_3$
  Perovskite Thin Films at Low Temperatures: Charge Transfer From the
  Orthorhombic Phase to the Tetragonal Phase}}.
\newblock \emph{\bibinfo{journal}{J. Phys. Chem. Lett.}}
  \textbf{\bibinfo{volume}{7}}, \bibinfo{pages}{2316} (\bibinfo{year}{2016}).

\bibitem{galkowski2016}
\bibinfo{author}{Galkowski, K.} \emph{et~al.}
\newblock \bibinfo{title}{{Determination of the exciton binding energy and
  effective masses for methylammonium and formamidinium lead tri-halide
  perovskite semiconductors}}.
\newblock \emph{\bibinfo{journal}{Energy Environ. Sci.,}}
  \textbf{\bibinfo{volume}{9}}, \bibinfo{pages}{962} (\bibinfo{year}{2016}).

\bibitem{zhang2014}
\bibinfo{author}{Zhang, X.}, \bibinfo{author}{Liu, Q.}, \bibinfo{author}{Luo,
  J.-W.}, \bibinfo{author}{Freeman, A.~J.} \& \bibinfo{author}{Zunger, A.}
\newblock \bibinfo{title}{Hidden spin polarization in inversion-symmetric bulk
  crystals}.
\newblock \emph{\bibinfo{journal}{Nat. Phys.}} \textbf{\bibinfo{volume}{10}},
  \bibinfo{pages}{387--393} (\bibinfo{year}{2014}).

\bibitem{druzbicki2016}
\bibinfo{author}{Dru\'zbicki, K.} \emph{et~al.}
\newblock \bibinfo{title}{{Unexpected Cation Dynamics in the Low-Temperature
  Phase of Methylammonium Lead Iodide: The Need for Improved Models}}.
\newblock \emph{\bibinfo{journal}{J. Phys. Chem. Lett.}}
  \textbf{\bibinfo{volume}{7}}, \bibinfo{pages}{4701--4709}
  (\bibinfo{year}{2016}).

\bibitem{beecher2016}
\bibinfo{author}{Beecher, A.~N.} \emph{et~al.}
\newblock \bibinfo{title}{{Direct Observation of Dynamic Symmetry Breaking
  above Room Temperature in Methylammonium Lead Iodide Perovskite}}.
\newblock \emph{\bibinfo{journal}{ACS Energy Lett.}}
  \textbf{\bibinfo{volume}{1}}, \bibinfo{pages}{880--887}
  (\bibinfo{year}{2016}).

\bibitem{niesner2016optical}
\bibinfo{author}{Niesner, D.} \emph{et~al.}
\newblock \bibinfo{title}{{Temperature-dependent optical spectra of
  single-crystal ${\mathrm{CH}}_{3}{\mathrm{NH}}_{3}{\mathrm{PbBr}}_{3}$
  cleaved in ultrahigh vacuum}}.
\newblock \emph{\bibinfo{journal}{Phys. Rev. B}} \textbf{\bibinfo{volume}{95}},
  \bibinfo{pages}{075207} (\bibinfo{year}{2017}).

\bibitem{kepenekian2015}
\bibinfo{author}{Kepenekian, M.} \emph{et~al.}
\newblock \bibinfo{title}{{Rashba and Dresselhaus Effects in Hybrid
  Organic--Inorganic Perovskites: From Basics to Devices}}.
\newblock \emph{\bibinfo{journal}{ACS Nano}} \textbf{\bibinfo{volume}{9}},
  \bibinfo{pages}{11557} (\bibinfo{year}{2015}).

\bibitem{Li2016}
\bibinfo{author}{Li, J.} \& \bibinfo{author}{Haney, P.~M.}
\newblock \bibinfo{title}{Optical spintronics in organic-inorganic perovskite
  photovoltaics}.
\newblock \emph{\bibinfo{journal}{Phys. Rev. B}} \textbf{\bibinfo{volume}{93}},
  \bibinfo{pages}{155432} (\bibinfo{year}{2016}).

\bibitem{zhang2017}
\bibinfo{author}{Zhang, C.}, \bibinfo{author}{Sun, D.} \&
  \bibinfo{author}{Vardeny, Z.~V.}
\newblock \bibinfo{title}{Multifunctional optoelectronic--spintronic device
  based on hybrid organometal trihalide perovskites}.
\newblock \emph{\bibinfo{journal}{Adv. Electron. Mater.}}
  \textbf{\bibinfo{volume}{3}} (\bibinfo{year}{2017}).

\bibitem{landolt2013}
\bibinfo{author}{Landolt, G.} \emph{et~al.}
\newblock \bibinfo{title}{{Bulk and surface Rashba splitting in single
  termination BiTeCl}}.
\newblock \emph{\bibinfo{journal}{New J. Phys.}} \textbf{\bibinfo{volume}{15}},
  \bibinfo{pages}{085022} (\bibinfo{year}{2013}).

\bibitem{ogawa2014}
\bibinfo{author}{Ogawa, N.}, \bibinfo{author}{Bahramy, M.},
  \bibinfo{author}{Kaneko, Y.} \& \bibinfo{author}{Tokura, Y.}
\newblock \bibinfo{title}{{Photocontrol of Dirac electrons in a bulk Rashba
  semiconductor}}.
\newblock \emph{\bibinfo{journal}{Phys. Rev. B}} \textbf{\bibinfo{volume}{90}},
  \bibinfo{pages}{125122} (\bibinfo{year}{2014}).

\bibitem{Ishizaka2011}
\bibinfo{author}{Ishizaka, K.} \emph{et~al.}
\newblock \bibinfo{title}{{Giant Rashba-type spin splitting in bulk BiTeI}}.
\newblock \emph{\bibinfo{journal}{Nat. Mater.}} \textbf{\bibinfo{volume}{10}},
  \bibinfo{pages}{521} (\bibinfo{year}{2011}).

\bibitem{liebmann2016}
\bibinfo{author}{Liebmann, M.} \emph{et~al.}
\newblock \bibinfo{title}{{Giant Rashba-Type Spin Splitting in Ferroelectric
  GeTe (111)}}.
\newblock \emph{\bibinfo{journal}{Adv. Mater.}} \textbf{\bibinfo{volume}{28}},
  \bibinfo{pages}{560--565} (\bibinfo{year}{2016}).

\bibitem{kim2014}
\bibinfo{author}{Kim, M.}, \bibinfo{author}{Im, J.}, \bibinfo{author}{Freeman,
  A.~J.}, \bibinfo{author}{Ihm, J.} \& \bibinfo{author}{Jin, H.}
\newblock \bibinfo{title}{{Switchable S= 1/2 and J= 1/2 Rashba bands in
  ferroelectric halide perovskites}}.
\newblock \emph{\bibinfo{journal}{Proc. Natl. Acad. Sci.}}
  \textbf{\bibinfo{volume}{111}}, \bibinfo{pages}{6900} (\bibinfo{year}{2014}).

\bibitem{leppert2016}
\bibinfo{author}{Leppert, L.}, \bibinfo{author}{Reyes-Lillo, S.~E.} \&
  \bibinfo{author}{Neaton, J.~B.}
\newblock \bibinfo{title}{{Electric Field-and Strain-Induced Rashba Effect in
  Hybrid Halide Perovskites}}.
\newblock \emph{\bibinfo{journal}{J. Phys. Chem. Lett.}}
  \textbf{\bibinfo{volume}{7}}, \bibinfo{pages}{3683--3689}
  (\bibinfo{year}{2016}).

\bibitem{kutes2014}
\bibinfo{author}{Kutes, Y.} \emph{et~al.}
\newblock \bibinfo{title}{{Direct Observation of Ferroelectric Domains in
  Solution-Processed CH$_3$NH$_3$PbI$_3$ Perovskite Thin Films}}.
\newblock \emph{\bibinfo{journal}{J. Phys. Chem. Lett.}}
  \textbf{\bibinfo{volume}{5}}, \bibinfo{pages}{3335} (\bibinfo{year}{2014}).

\bibitem{zhang2015magnet}
\bibinfo{author}{Zhang, C.} \emph{et~al.}
\newblock \bibinfo{title}{Magnetic field effects in hybrid perovskite devices}.
\newblock \emph{\bibinfo{journal}{Nat. Phys.}} \textbf{\bibinfo{volume}{11}},
  \bibinfo{pages}{427--434} (\bibinfo{year}{2015}).

\bibitem{guo2015spatial}
\bibinfo{author}{Guo, Z.}, \bibinfo{author}{Manser, J.~S.},
  \bibinfo{author}{Wan, Y.}, \bibinfo{author}{Kamat, P.~V.} \&
  \bibinfo{author}{Huang, L.}
\newblock \bibinfo{title}{{Spatial and temporal imaging of long-range charge
  transport in perovskite thin films by ultrafast microscopy}}.
\newblock \emph{\bibinfo{journal}{Nat. Commun.}} \textbf{\bibinfo{volume}{6}},
  \bibinfo{pages}{7471} (\bibinfo{year}{2015}).

\bibitem{shi2015}
\bibinfo{author}{Shi, D.} \emph{et~al.}
\newblock \bibinfo{title}{Low trap-state density and long carrier diffusion in
  organolead trihalide perovskite single crystals}.
\newblock \emph{\bibinfo{journal}{Science}} \textbf{\bibinfo{volume}{347}},
  \bibinfo{pages}{519--522} (\bibinfo{year}{2015}).

\bibitem{zhou2016giant}
\bibinfo{author}{Zhou, Y.} \emph{et~al.}
\newblock \bibinfo{title}{{Giant photostriction in organic-inorganic lead
  halide perovskites}}.
\newblock \emph{\bibinfo{journal}{Nat. Commun.}} \textbf{\bibinfo{volume}{7}},
  \bibinfo{pages}{11193} (\bibinfo{year}{2016}).

\bibitem{gottesman2014extremely}
\bibinfo{author}{Gottesman, R.} \emph{et~al.}
\newblock \bibinfo{title}{{Extremely slow photoconductivity response of
  CH$_3$NH$_3$PbI$_3$ perovskites suggesting structural changes under working
  conditions}}.
\newblock \emph{\bibinfo{journal}{J. Phys. Chem. Lett.}}
  \textbf{\bibinfo{volume}{5}}, \bibinfo{pages}{2662--2669}
  (\bibinfo{year}{2014}).

\bibitem{gottesman2015photoinduced}
\bibinfo{author}{Gottesman, R.} \emph{et~al.}
\newblock \bibinfo{title}{{Photoinduced reversible structural transformations
  in free-standing CH$_3$NH$_3$PbI$_3$ perovskite films}}.
\newblock \emph{\bibinfo{journal}{J. Phys. Chem. Lett.}}
  \textbf{\bibinfo{volume}{6}}, \bibinfo{pages}{2332--2338}
  (\bibinfo{year}{2015}).

\end{thebibliography}

\end{document}